\RequirePackage{fix-cm}
\documentclass{svjour3}                     % onecolumn (standard format)

\smartqed  % flush right qed marks, e.g. at end of proof
\usepackage{graphicx}
\usepackage{hyperref,overpic}

\hypersetup{
	colorlinks=true,
	linktoc=all,
	linkcolor=blue,
	citecolor=blue,
}
\usepackage{array}
\newcolumntype{M}[1]{>{\centering\arraybackslash}m{#1}}

\usepackage{amsmath,amssymb}

\usepackage{color}
\definecolor{red}{rgb}{0.8500, 0.1250, 0.0480} % added
\definecolor{blue}{rgb}{0, 0.4470, 0.7410}
\definecolor{green}{rgb}{0.4660, 0.6740, 0.1880}
\definecolor{gray}{rgb}{0.7, 0.7, 0.7}
\begin{document}

\title{Spanwise effects on instabilities of compressible flow over a long rectangular cavity
\thanks{This work was supported by the US Air Force Office of Scientific Research (Grant number: FA9550-13-1-0091; Program Managers: Douglas Smith and Ivett Layva).}}

\author{Y. Sun         \and
        	    K. Taira     \and
	    L. N. Cattafesta III       \and
	    L. S. Ukeiley %etc.
}

\institute{Y. Sun \and K. Taira \and L. N. Cattafesta\at
              Department of Mechanical Engineering, Florida State University \\
              \email{ktaira@fsu.edu}           \\ \\
	      L. S. Ukeiley \at
	      Department of Mechanical and Aerospace Engineering, University of Florida
}

\date{Received: date / Accepted: date}

\maketitle

\begin{abstract}
The stability properties of two- (2D) and three-dimensional (3D) compressible flows over a rectangular cavity with length-to-depth ratio of $L/D=6$ is analyzed at a free stream Mach number of $M_\infty=0.6$ and depth-based Reynolds number of $Re_D=502$. In this study, we closely examine the influence of three-dimensionality on the wake-mode that has been reported to exhibit high-amplitude fluctuations from the formation and ejection of large-scale spanwise vortices.  Direct numerical simulation (DNS) and bi-global stability analysis are utilized to study the instability characteristics of the wake-mode.  Using the bi-global stability analysis with the time-average flow as the base state, we capture the global stability properties of the wake-mode at a spanwise wavenumber of $\beta=0$. To uncover spanwise effects on the 2D wake-mode, 3D DNS are performed with cavity width-to-depth ratio of $W/D=1$ and $2$. We find that the 2D wake-mode is not present in the 3D cavity flow for a wider spanwise setting with $W/D=2$, in which spanwise structures are observed near the rear region of the cavity. These 3D instabilities are further investigated via bi-global stability analysis for spanwise wavelengths of $\lambda/D=0.5-2.0$ to reveal the eigenspectra of the 3D eigenmodes. Based on the findings of 2D and 3D global stability analysis, we conclude that the absence of the wake-mode in 3D rectangular cavity flows is due to the release of kinetic energy from the spanwise vortices to the streamwise vortical structures that develops from the spanwise instabilities.
  

\end{abstract}

\section{Introduction}
\label{sec:intro}

Cavity flow is a fundamental research topic with numerous applications.  A recent review by Lawson and Barakos \cite{Lawson:PAS11} summarizes the progress of experimental and computational work on cavity flows during the past decades, including fundamental flow physics and advanced techniques on analyzing the flows. Controls on altering undesirable cavity flow features are reviewed by Colonius \cite{Colonius:AIAA01} and Cattafesta et al.~\cite{cattafesta2003review}, with focus on flow control performance and current challenges in implementation. Findings from global stability analysis \cite{Bres:JFM08} are also leveraged to suppress intense fluctuations in open-cavity flows \cite{George:AIAA15,Zhang:AIAA15}. 

In open-cavity flow, a shear layer spanning over the cavity amplifies free stream disturbances due to the Kelvin--Helmholtz instability. When the shear layer impinges on the trailing edge of the cavity, intense pressure fluctuations are generated and propagate upstream to introduce additional perturbations.  This natural feedback process associated with the oscillations in the flow forms cavity flow resonance, which is referred to as the {\it shear layer mode} or {\it Rossiter mode} followed by the earliest description and semi-empirical formula given by Rossiter \cite{Rossiter:ARCRM64}.  The influence of Reynolds number, Mach number, cavity aspect ratio, and incoming boundary layer profile on the shear layer instability has been studied in detail in the past \cite{Krishnamurty:1956,Rockwell:JFE78,Colonius:AIAA01,Rowley:JFM02}.

Besides the shear layer mode in open-cavity flow, there is another more violent oscillation mechanism named as {\it wake-mode} reported in a few experimental  \cite{Gharib:JFM87,Zhang:AIAA06,Zhang:EF11} and numerical studies \cite{Rowley:JFM02,Bres:2007}. It was firstly reported in the experimental work by Gharib and Roshko \cite{Gharib:JFM87} that the wake-mode appears in an axisymmetric incompressible cavity flow.  Large vortices are formed at the cavity leading edge and impinge on the trailing edge inducing more intense fluctuations than those of shear layer modes.  The wake-mode was also observed in numerical simulations of compressible flow over long rectangular cavities (with length-to-depth ratio of $L/D=4$) conducted by Rowley et al.~\cite{Rowley:JFM02} and Br\`es \cite{Bres:2007}. They reported that the wake-mode is a hydrodynamic instability mode because its oscillation frequency is independent of free stream Mach number. However, the wake-mode has not been observed in experiments of flows over a rectangular cavity, which makes the existence of the wake-mode an open question.

In order to uncover the stability characteristics of open-cavity flows, besides the experimental and nonlinear numerical studies mentioned above, global linear stability theory has also been utilized and become attractive because of the rich instability properties that can be captured. In the scope of linear stability analysis for open-cavity flow, the framework of bi-global stability analysis by solving an eigenvalue problem around a 2D base state was reported by Theofilis and Colonius \cite{Theofilis:AIAA04}, in which they examined 3D eigenmodes of compressible open-cavity flow with $M_\infty=0.325$ and $L/D=2$. Following this work, Br\`es and Colonius \cite{Bres:JFM08} further investigated  dominant 3D global modes of cavity flow via performing linear simulations for an extensive parametric study. On evaluating linear stability analysis results obtained from numerical approach, Vicente et al.~\cite{Vicente:JFM14} examined the three-dimensionality of incompressible open-cavity flow with $L/D=2$ in experiments. They found the spatial structures obtained from linear stability analysis are almost consistent to those in saturated flow fields from the experiments, but the spanwise boundary conditions can affect the properties of the 3D modes captured from linear stability analysis. While several stability analyses have been performed \cite{Theofilis:AIAA04,Bres:JFM08,Yamouni:JFM13,Vicente:JFM14,Citro:JFM15}, the stability characteristics of the wake-mode and the effects of spanwise instabilities on the wake-mode has not been examined in detail.

Moreover, there are a number of studies \cite{Maull:JFM63,Beresh:AIAA15,George:AIAA15,Arunajatesan:AIAA14,Sun:AIAA16} performed on examining the influence of cavity width and sidewalls which are known to modify the flow field greatly. In recent experiments, the work by Beresh et al.~\cite{Beresh:AIAA15} on supersonic flow over a rectangular cavity showed different mean flow patterns as a function of $W/D$. The study by George et al.~\cite{George:AIAA15} on cavity flows with full span and finite span also showed the variation of the mean flow due to the cavity span configuration. In numerical studies, cavity width and sidewall effects on flow fields were also examined by Arunajatesan et al.~\cite{Arunajatesan:AIAA14} and Sun et al.~\cite{Sun:AIAA16}.

In this study, we perform bi-global stability analysis of flow over a long rectangular cavity of $L/D = 6$ and compare our findings with 2D and 3D DNS.  In Section \ref{sec:approach}, we present the numerical approach taken to perform direct numerical simulations and bi-global stability analysis of compressible open-cavity flows.  Based on this formulation, in Section \ref{sec:results} we discuss the characteristics of the wake-mode in 2D and 3D cavity flows obtained from simulations, and eigen properties of 2D and 3D modes uncovered via the bi-global stability analysis. At last, we provide Section \ref{sec:summary} to summarize our findings that the wake-mode is a 2D instability and is hardly observed in 3D settings due to the appearance of spanwise instability modes.

\section{Analysis approach}
\label{sec:approach}

In the present study, we first carry out 2D DNS to examine the nonlinear open-cavity flows with a wake-mode, which generates high-amplitude fluctuations.  Bi-global stability analysis is then conducted with the 2D base state and spanwise wavenumber of $\beta=0$ to uncover 2D eigenmode characteristics that can be related to the wake-mode captured in 2D DNS. Second, spanwise effects on the wake-mode are studied by performing 3D DNS.  For a sufficiently wide cavity, the wake-mode is not prominent. Spanwise motion observed in the 3D flow fields is further investigated via the bi-global stability analysis to capture the 3D eigenmodes. Finally, spanwise effects on the three-dimensional flow are discussed to explain the suppression of the 2D wake-mode in the 3D flow fields.    

\begin{figure}[hbpt]
\begin{center}
  \includegraphics[width=0.8\textwidth]{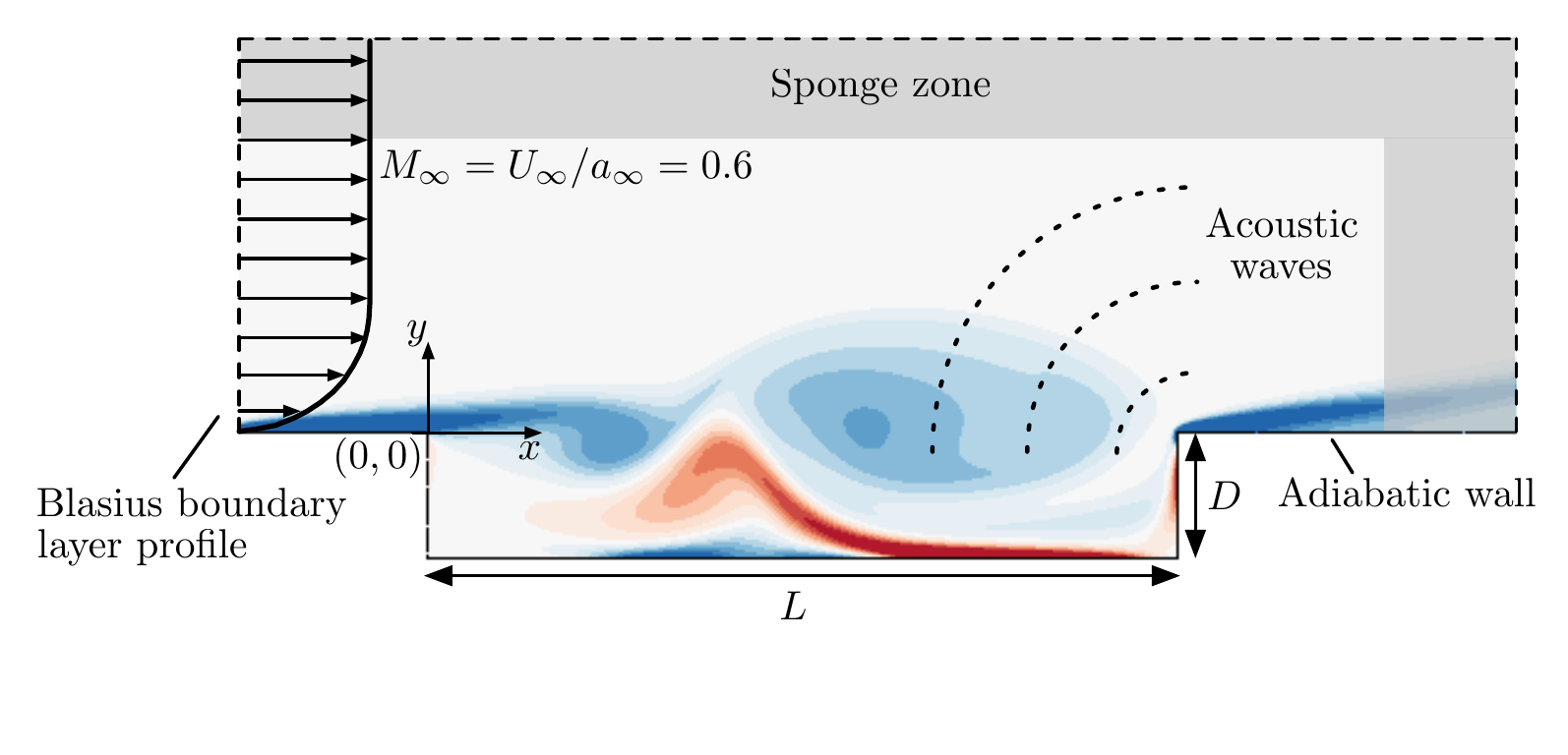}
  \vspace{-5mm}
\caption{The computational setup for cavity with $L/D=6$ (not to scale). Shown in the background is the instantaneous vorticity $\omega_z D/U_\infty$.}
\label{fig:setup} 
\end{center}     
\end{figure}

We perform 2D and 3D DNS with a high-fidelity compressible flow solver CharLES \cite{Bres:AIAA2012} to characterize the stability of open-cavity flows. The cavity aspect ratio is set to $L/D=6$ with incoming flow at $M_\infty=0.6$ and $Re_D=U_\infty D/\nu=502$ as shown in Fig. \ref{fig:setup}. The ratio of cavity depth and initial momentum boundary layer thickness at the leading edge of cavity is $D/\theta_0=26.4$. The flow is set to be periodic in the spanwise direction. Additional details on the computational setup can be found in our previous work \cite{Sun:AIAA2014,Sun:AIAA16}. The regions (farfield and outflow) where sponge zones are applied have been chopped out when we prepare the base states. Only physical solutions are used in the biglobal stability analysis.

Bi-global stability analysis \cite{Theofilis:PAS03,Theofilis:ARFM11} is performed with respect to a 2D base state for a given spanwise wavenumber $\beta$ (wavelength: $\lambda=2\pi/\beta$), to determine the dominant eigenmodes and their eigenvalues of compressible flow over a rectangular cavity.  We follow the standard formulation \cite{Schmid01} and decompose the conservative state variables $q(x,y,z,t)=[\rho, \rho u, \rho v, \rho w, e]^T$ into a 2D base state $\bar q(x,y)$ and a perturbation $q'(x,y,z,t)$. By substituting $q=\bar q +q'$ into the compressible Navier--Stokes (NS) equations and retaining the linear terms, we find the linearized NS equations
\begin{equation}
\frac{\partial q'}{\partial t}=L(\bar{q})q',
\label{lin}
\end{equation}  
where $L$ is a linear operator.  With this formulation, a general modal formulation can be assumed for the perturbation
\begin{equation}
q'(x,y,z,t)=\hat q(x,y) e^{i({\beta z - \omega t})}+c.c.,
\end{equation}
where $\hat q$ is its amplitude function (eigenvector) and $c.c.$ represents its complex conjugate.  Consequently, the linearized governing equation, Eq. (\ref{lin}), can be transformed to an eigenvalue problem
\begin{equation}
-i\omega \hat q=A(\bar q; \beta)\hat q.
\label{eigen}
\end{equation}
Here, the eigenvalue $\omega=\omega_r+i \omega_i$ is a complex number. The real component $\omega_r$ represents the frequency of the eigenmode, which in this paper is normalized as Strouhal number $St_D=\omega_r D/2\pi U_\infty$. The imaginary component $\omega_i$ corresponds to the growth ($\omega_i>0$) or decay ($\omega_i<0$) rate of corresponding eigenmode. ARPACK \cite{Arpack:96} is used to solve the above large-scale eigenvalue problem, Eq. (\ref{eigen}). The present approach is based on matrix-free computation. Since the linear governing equations for the perturbation variables are explicit in their forms, we use the regular mode (not shift-and-invert transformation) in ARPACK and only provide values of the matrix vector products $A(\bar q; \beta)\hat q$ repeatly to the solver, which avoids requesting large memory space to store matrix entries while solving the eigenvalue problem. All the eigenmodes reported in this paper are converged with $||-i\omega \hat q-A\hat q||\le\mathcal{O}(10^{-10})$.

In the present work, we consider the use of both the mean flow and the unstable steady state as the base flow in the bi-global stability analysis.  For the unstable steady state, the selective frequency damping method \cite{Akervik:PF06} is utilized because the flow in this study is inherently unstable.  In the linear bi-global stability analysis below, 2D and 3D eigenmodes are examined with spanwise wavelengths of $0 \le \lambda/D \le 2$. Along the cavity walls, velocity perturbations and the wall-normal pressure are set to zero. According to the integral form of the governing equations, boundary condition for density perturbation is not required because the momentum perturbation flux is zero due to base velocity along the wall. For the inlet, density and velocity perturbations, as well as the pressure gradient are prescribed to be zero. For the outflow and farfield boundaries, Neumann boundary condition of density, velocity and pressure are prescribed as zero. Moreover, adiabatic conditions is assumed for all boundaries. A finite volume method with second-order spatial accuracy is utilized. The obtained base states from 2D DNS with 0.5 million grid points are interpolated onto a coarse mesh with 60,400 non-uniform grid points. A refined mesh with 134,400 non-uniform grid points is used as well for the eigenvalue problem and the eigen-spectra overlap with only 0.059\% difference, thus, we select the mesh with 60,400 for the present study.

\begin{figure}[hbpt]
\begin{center}
  \includegraphics[width=0.98\textwidth]{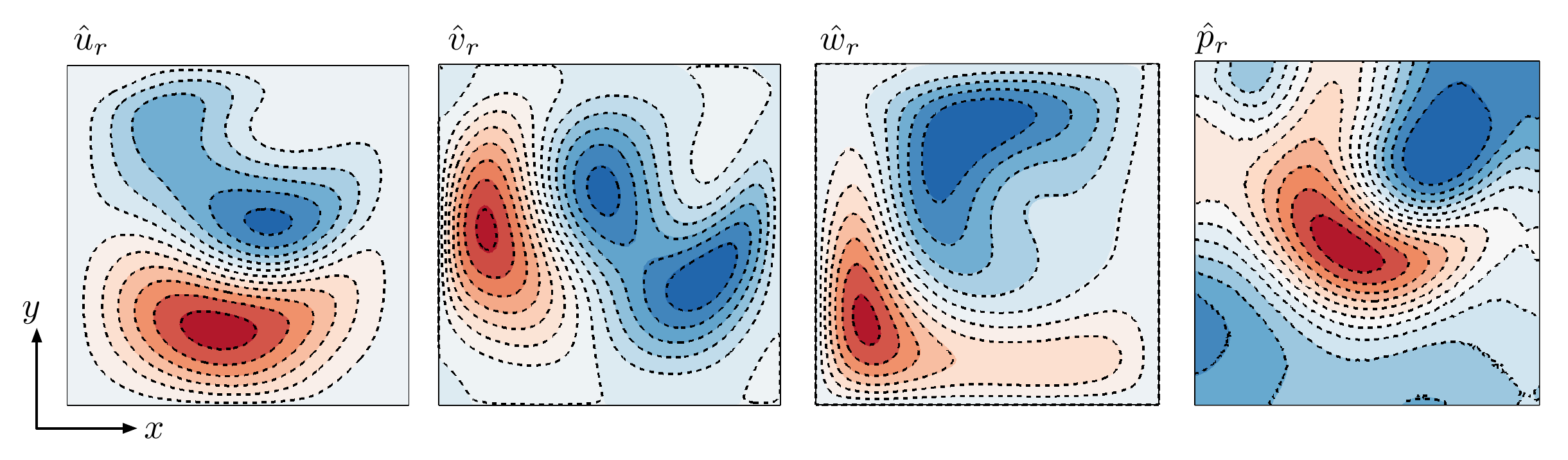}
\caption{Comparison between the present results (contours) and those from an incompressible stability analysis code \cite{Munday:inhouse16} (dashed lines) performed for validation.  Shown are the real component of velocity $\hat u,\hat v,\hat w$ and pressure $\hat p$ of the least-stable eigenmode of the lid-driven cavity flow at $Re=200$ with $\beta=1$.}
\label{fig:ldc} 
\end{center}     
\end{figure}

We also compare our results from an incompressible lid-driven cavity flow ($L/D=1$) and a compressible open-cavity flow ($L/D=2$) with those reported in the literature \cite{Ramanan:PF94,Ding:JCP98,Theofilis:JFM04}. The dominant eigenvalue and eigenvector are compared to those in previous work in Table \ref{tab:validation} and Fig. \ref{fig:ldc} and are found to be in good agreement for both flows, providing confidence in the present bi-global stability analysis formulation.

\begin{table}[hbpt]
\centering
\caption{Comparison of the computed growth/decay rate $\omega_i$ of the most-unstable/least-stable eigenmode for lid-driven and open-cavity flows.  For the lid-driven cavity flow: $Re=200$ and $\beta = 1$ (present: $M_\text{lid}=0.025$, all others: incompressible). For the open-cavity flow: $Re_D=1500$ (present: $M_\infty = 0.3$ and Br\`es \& Colonius \cite{Bres:JFM08}: $M_\infty=0.325$.)}
\label{tab:validation}       
\begin{tabular}{lc|cccccc}
\noalign{\smallskip}
\multicolumn{2}{l}{Lid-driven cavity flow}				&\multicolumn{3}{l}{Open-cavity flow	}\\ \hline \hline
	 								& $(\omega_i) i$ 	&				&\multicolumn{2}{c}{$St_D+ (\omega_i D/U_\infty)i$}	\\ 
									&			&				& \\
Ramanan \& Homsy \cite{Ramanan:PF94} 	& $-0.34i$ 	&$\lambda/D$		&Br\`es \& Colonius \cite{Bres:JFM08} 	& Present\\
Ding \& Kawahara \cite{Ding:JCP98}  		& $-0.3183i$  	&0.5				&$0.0000 - 0.0180i$	&$0.0000-0.0136i$		\\
Theofilis et al.~\cite{Theofilis:JFM04}  		& $-0.3297i $	&1.0				&$0.0248 + 0.0086i$	&$0.0248+0.0085i$		\\
Munday \& Taira \cite{Munday:inhouse16}&$ -0.3298i$  	&1.5			&$0.0104 - 0.0019i$	&$0.0108-0.0022i$		\\
Present								& $-0.3297i$ 	&2.0				&$0.0130 - 0.0052i$	&$0.0138-0.0056i$		\\
\hline
\end{tabular}
\end{table}

{\color{black}
\section{Results}
\label{sec:results}

\subsection{2D global stability}

\subsubsection{2D base states} 

For the flow condition considered in the present work, the wake-mode is observed in the 2D simulation. As visualized in Fig. \ref{fig:spec}, a large vortex rolls up at the center of the cavity and sheds from upstream at approximately 2.5-cavity depth away from the cavity leading edge, as an opposite sign vortex sheet emerges from the floor. When the large vortex approaches the trailing edge, an opposite sign vortex sheet is squeezed in between the large vortex and the second vortex in formation from the leading edge.  As the large vortex interacts with the trailing edge, it induces large fluctuations and advects downstream, and a new vortex continues to grow in the center of the cavity. By analyzing the time history of velocity extracted at $(L/2,0)$, using Fourier transform, the wake-mode frequency $St_D=fD/U_\infty=0.0572$ and its harmonics can be identified as shown in Fig. \ref{fig:spec}, which agrees with the studies by Rowley et al.~\cite{Rowley:JFM02}. 

\begin{figure}[hbpt]
\begin{center}
  \includegraphics[width=4.8in]{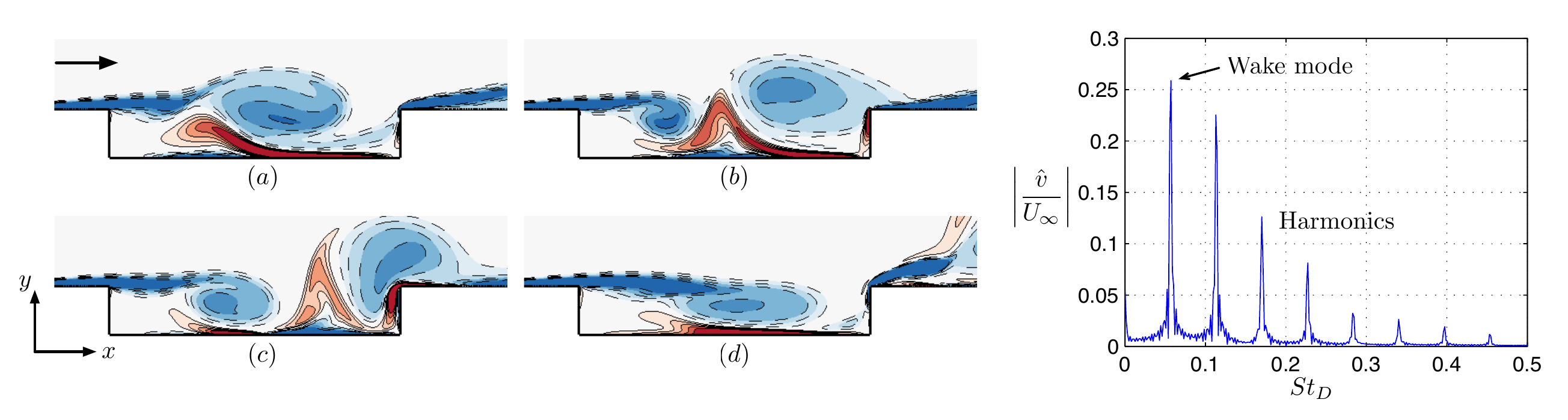}
\caption{Instantaneous vorticity fields $\omega_z D/U_\infty\in(-2.4,2.4)$ are shown in subplots $(a)$ - $(d)$ for every quarter period of the wake-mode.  Also presented is the spectrum of normal velocity $v/U_\infty$ at the center of the cavity $(L/2, 0)$.}
\label{fig:spec} 
\end{center}     
\end{figure}

In order to perform the bi-global stability analysis, a 2D base state needs to be prescribed.  In this investigation, we determine the unstable steady state and mean flow as shown in Fig. \ref{fig:base}.  The unstable steady state is computed by utilizing the selective frequency damping method \cite{Akervik:PF06}, while the mean flow is determined by temporally averaging the unsteady flow field.  The streamlines of these two base states are quite different in terms of the locations and shape of the major recirculation zones.  For the unstable steady state, the center of rotation sits towards the rear part of the cavity. However, for the mean flow, there are two recirculation regions with the center of the major recirculation being located near the center of the cavity.  In our companion work \cite{Sun:JFM16} on examining Mach number ($0.1\le M_\infty \le1.4$) effects on the flow for the same configuration as here, we find the streamlines of mean flows at other Mach numbers considered (which exhibit shear layer instabilities) are similar to the unstable steady state in Fig. \ref{fig:base} $(a)$. Hence, one can postulate the existence of the shear layer mode or the wake-mode in the flow field attributes to the difference in the mean flow profile.  

\begin{figure}[hbpt]
\begin{center}
  \includegraphics[width=4.8in]{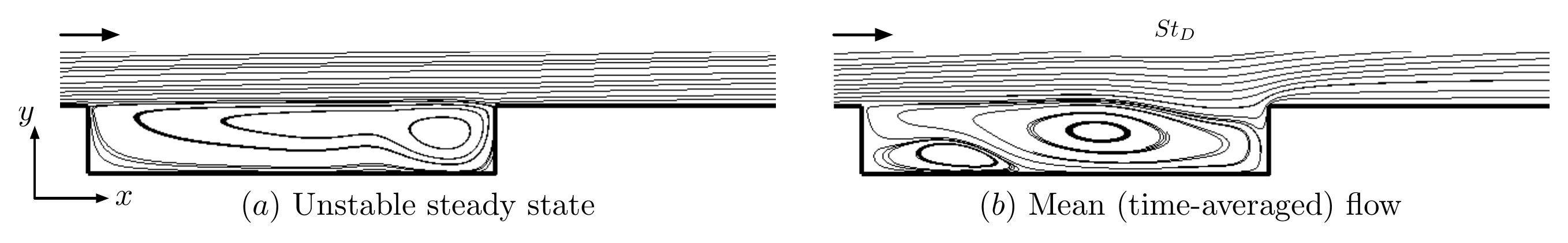}
\caption{Streamlines of base states from 2D DNS: $(a)$ unstable steady state and $(b)$ mean (time-average) flow. }
\label{fig:base} 
\end{center}     
\end{figure}

\subsubsection{2D eigenmodes for $\lambda/D=\infty$ $(\beta=0)$ }

The linear bi-global analysis with respect to the mean flow and the unstable steady state discussed above is presented here.  The bi-global stability analysis is performed first with $\beta=0$ to obtain the 2D eigenmodes. The dominant and subdominant eigenmodes extracted from the analysis are listed in Table \ref{tab:2Deigenvalue}.  For the stability of flow about the unstable steady state, Rossiter modes I - IV are identified from the eigenspectrum.  Among the four captured Rossiter modes, only mode II and III are unstable.  In the case of using the mean flow as the base state, wake-mode w-i is uncovered with $St_D=0.0551$ which is in agreement with the oscillatory frequency $St_D=0.0572$ from the 2D nonlinear simulation with only 4\% difference.  Using the mean flow as the base for the stability analysis, we find that the two eigenmodes w-ii and w-iii with higher frequencies are approximately the harmonics of the dominant wake-mode w-i, and the only unstable mode is mode w-ii with $St_D=0.1218$.

\begin{table}[hbpt]
\caption{Leading eigenvalues from the bi-global stability analysis using the unstable steady state and the mean flow as the base state. Also listed is the Strouhal number from the 2D nonlinear simulation. }
\begin{center}
{\scriptsize
\begin{tabular}{cc|cc|cc}
\multicolumn{2}{c}{Unstable steady state}		& \multicolumn{2}{c}{Mean flow} 	&2D nonlinear simulation\\ \hline
Rossiter modes 	& $St_D+(\omega_i D/U_\infty)i$ 	& wake-modes 	& $St_D+(\omega_i D/U_\infty)i$	&$St_D$\\ 
				&							&										&		 \\
I			&$0.0636 -  0.0002i$				&w-i			&$0.0551 -  0.0074i$					&0.0572	\\
II 			&$0.1307 + 0.0808i$				&w-ii			&$0.1218 + 0.0085i$							&		\\
III			&$0.1959 + 0.0685i$				&w-iii		&$0.1859 -  0.0327i$							&		\\
IV			&$0.2624 -  0.0049i$				&			&							&		\\ \hline
\end{tabular}
}
\end{center}
\label{tab:2Deigenvalue}
\end{table}

The spatial structures of the eigenmodes obtained by using two different base states are compared in Fig.~\ref{fig:uss_eigenvector}.   As it can be seen from the contours of Rossiter mode eigenvectors, higher frequency modes have smaller structure size.  The largest spatial fluctuations of Rossiter modes are spatially located along the shear layer, aligned horizontally with the cavity surface at the trailing edge, which is consistent with the profile of the base state as seen in Fig. \ref{fig:base} (a). In the case of the wake-modes, the dominant spatial oscillation of the eigenvector deflects upwards near the trailing edge, which is caused by the abrupt change in the curvature of the time-average streamlines near the trailing edge, as shown in Fig. \ref{fig:base} (b).

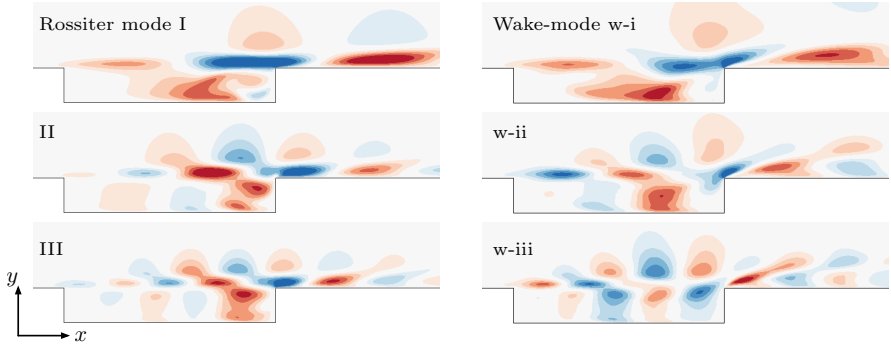
\begin{figure}[hbpt]
\begin{center}
\begin{overpic}[width=4.7in]{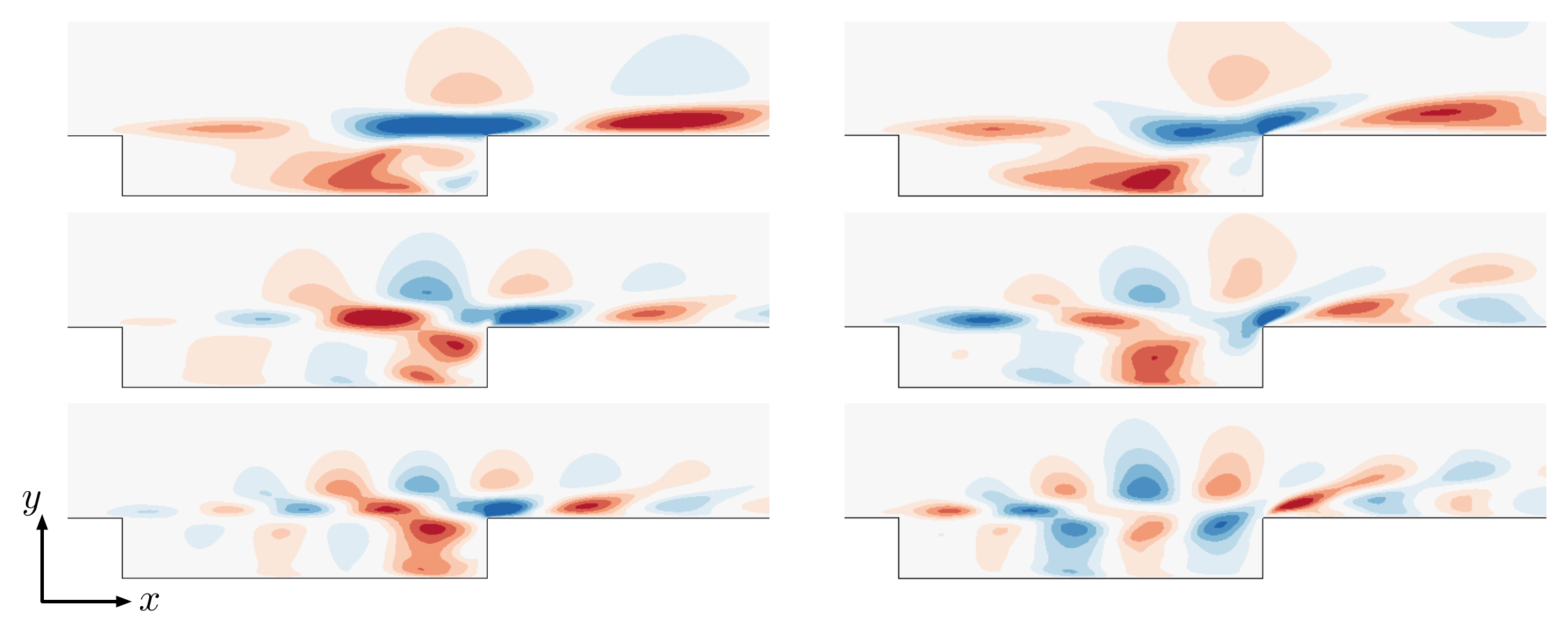}
\put(5,36){\scriptsize Rossiter mode I}
\put(5,24){\scriptsize II}
\put(5,11){\scriptsize III}
\put(55,36){\scriptsize Wake-mode w-i}
\put(55,24){\scriptsize w-ii}
\put(55,11){\scriptsize w-iii}
\end{overpic}
\caption{Contours of the real component $\hat u_r$ of the eigenvector for the Rossiter modes (left) and the wake-modes (right) based upon the unstable steady state and the mean flow, respectively. The contour levels of $\hat u_r/U_\infty \in[-0.008,0.008]$ are used.}
\label{fig:uss_eigenvector} 
\end{center}
\end{figure}

We should note that Fig. \ref{fig:uss_eigenvector} only represents the eigenvector at a certain phase.  As shown in Fig. \ref{fig:rotate_vec}, we compare the spatial structures of spanwise vorticity $\omega_z D/U_\infty$ at other phases of the Rossiter modes  and the wake-modes.  We find that for the Rossiter modes, the disturbances inside the cavity and those in the shear layer do not exhibit much interaction.  Moreover, we observe the disturbance recirculating at the rear part of the cavity where a recirculation zone appears in the unstable steady state.  On the other hand, for the wake-modes, the disturbances inside the cavity continuously penetrate into the shear layer region at the trailing edge. Additionally, the disturbances inside the cavity are active around the center of the cavity where the recirculation is located in the mean flow, as shown in Fig. \ref{fig:base} (b). Even though the eigenvector features of the Rossiter mode and the wake-mode differ inside the cavity and around the trailing edge, they behave in a similar manner in the fore region of the cavity where these two types of base states remain similar.

\begin{figure}[hbpt]
\begin{center}
{\scriptsize
\renewcommand{\arraystretch}{0.8}
\begin{tabular}{M{0.8cm}M{4.5cm}M{4.5cm}}
$\varphi-\varphi_0$&Rossiter mode I& Wake-mode w-i		\\ \hline 
$\pi/5$ 		&			   \includegraphics[width=0.4\textwidth]{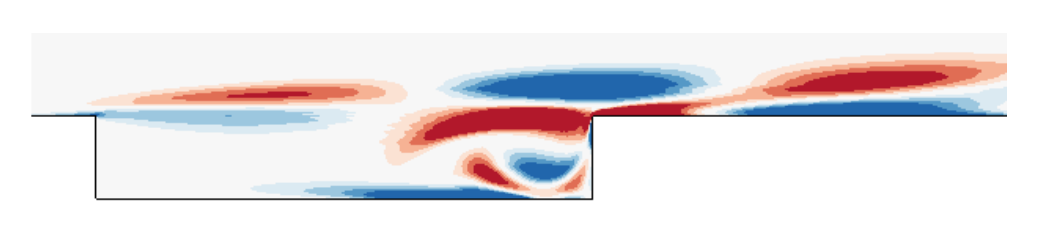}	& \includegraphics[width=0.4\textwidth]{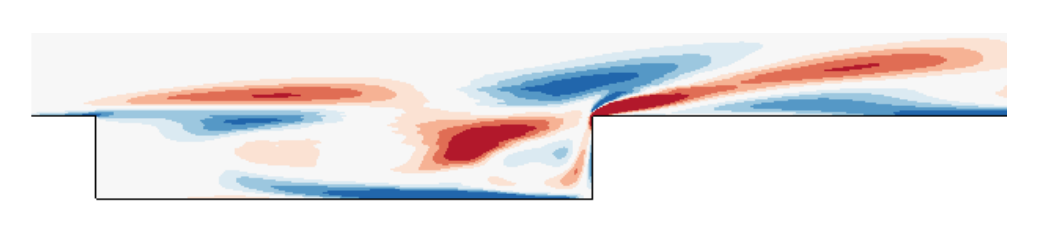} \\ 
$3\pi/5$  		&\vspace{-0.1in}  \includegraphics[width=0.4\textwidth]{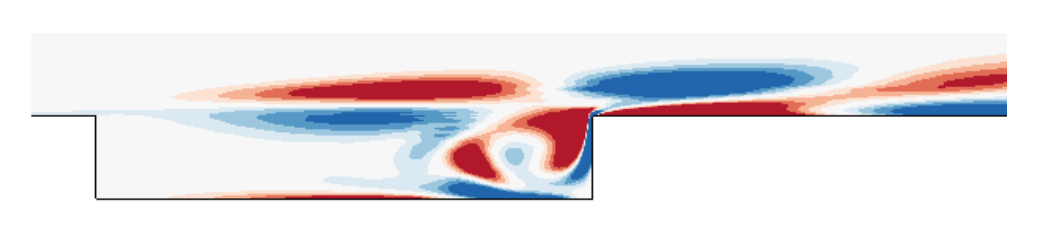}	&\vspace{-0.1in}\includegraphics[width=0.4\textwidth]{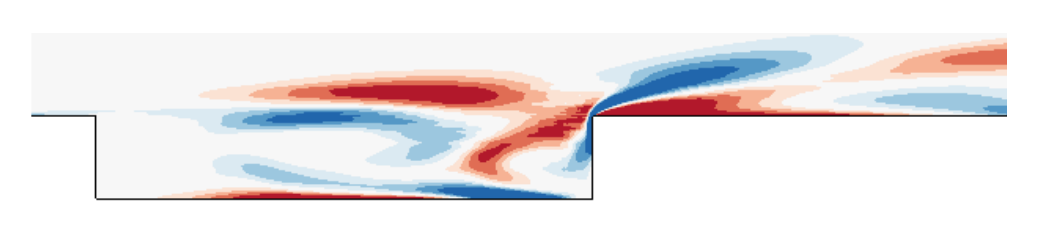}\\
$\pi$  		&\vspace{-0.1in}  \includegraphics[width=0.4\textwidth]{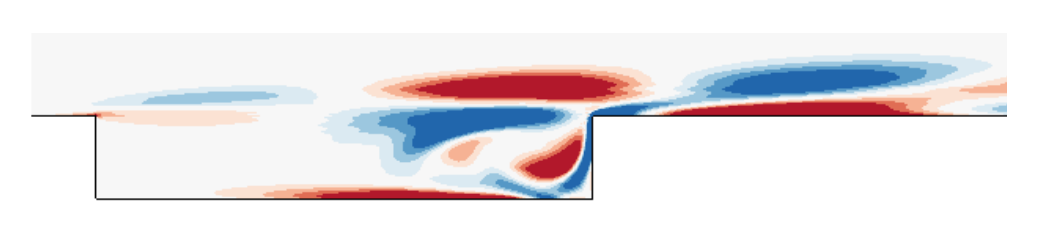}	&\vspace{-0.1in}\includegraphics[width=0.4\textwidth]{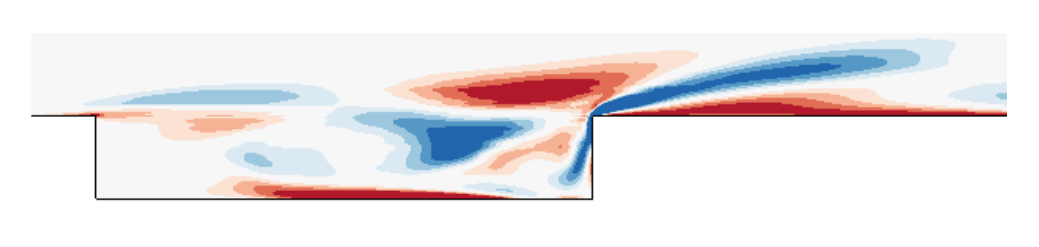}\\ 
			&Rossiter mode II												&Wake-mode w-ii		\\ \hline
$\pi/5$ 		&  			   \includegraphics[width=0.4\textwidth]{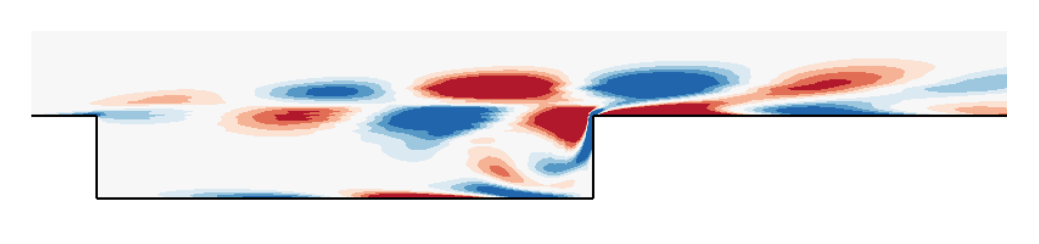}	& \includegraphics[width=0.4\textwidth]{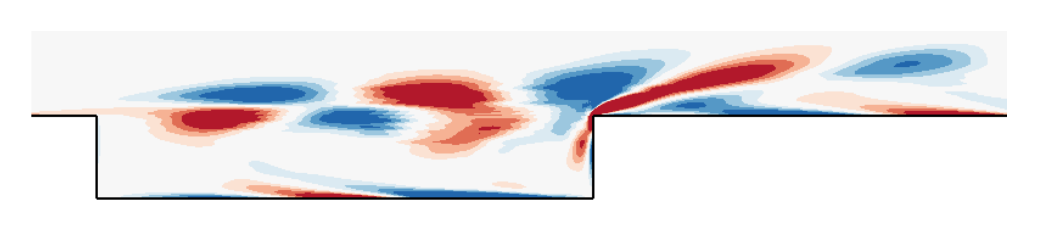}\\ 
$3\pi/5$  		&\vspace{-0.1in}  \includegraphics[width=0.4\textwidth]{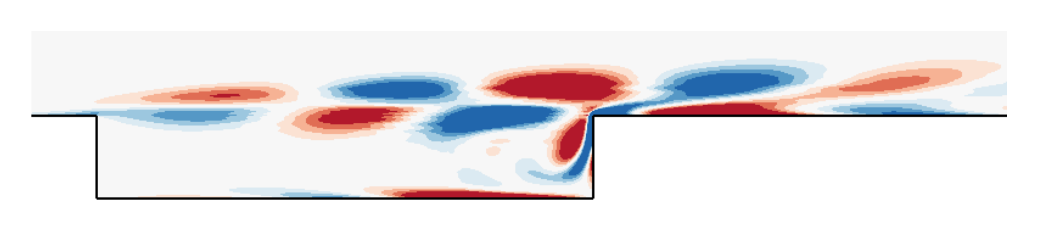}	&\vspace{-0.1in}\includegraphics[width=0.4\textwidth]{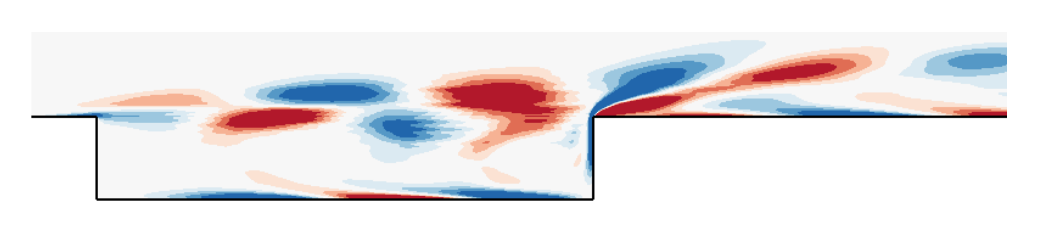}\\
$\pi$  		&\vspace{-0.1in}  \includegraphics[width=0.4\textwidth]{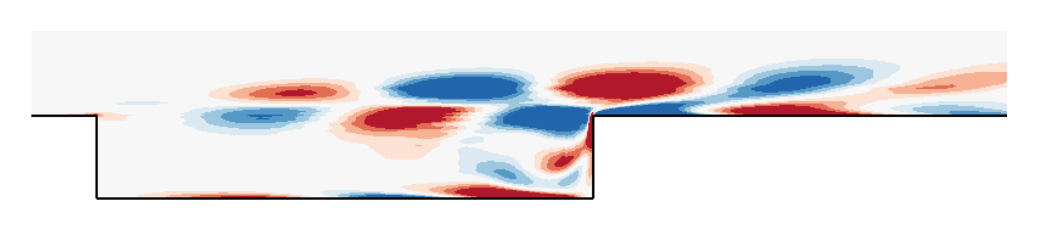}	&\vspace{-0.1in}\includegraphics[width=0.4\textwidth]{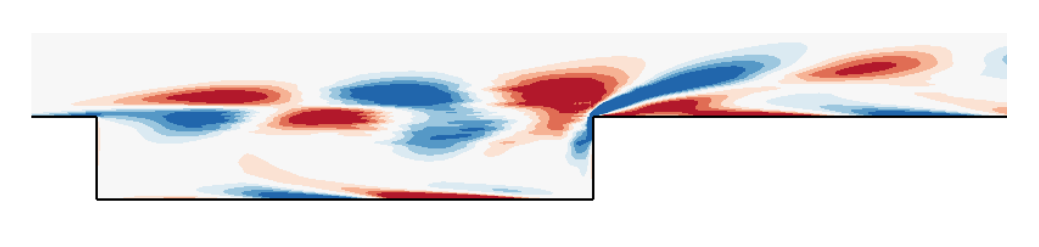}\\
			&Rossiter mode III												&Wake-mode w-iii		\\ \hline
$\pi/5$ 		& 			   \includegraphics[width=0.4\textwidth]{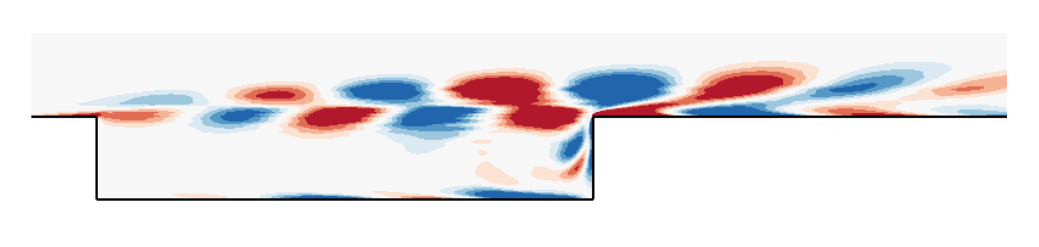}	&\includegraphics[width=0.4\textwidth]{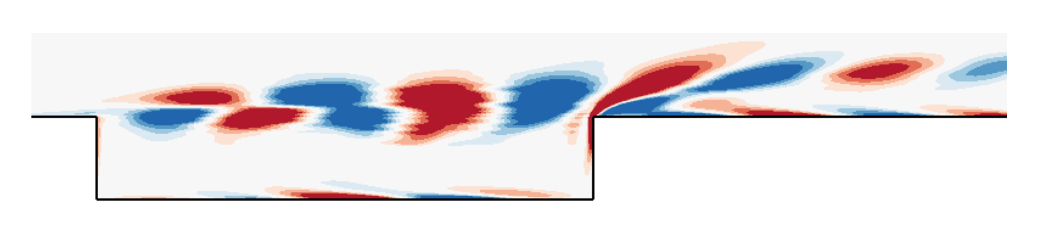}\\ 
$3\pi/5$  		&\vspace{-0.1in}  \includegraphics[width=0.4\textwidth]{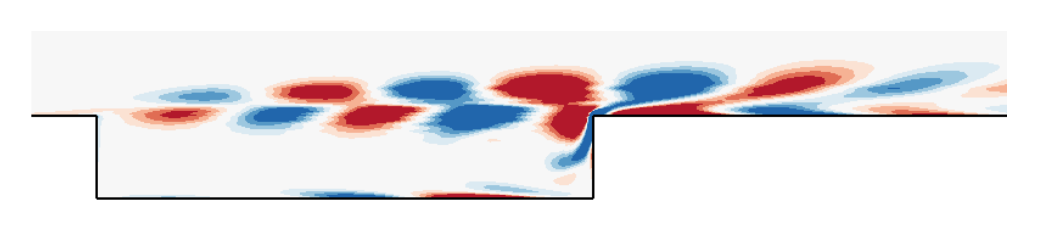}	&\vspace{-0.1in}\includegraphics[width=0.4\textwidth]{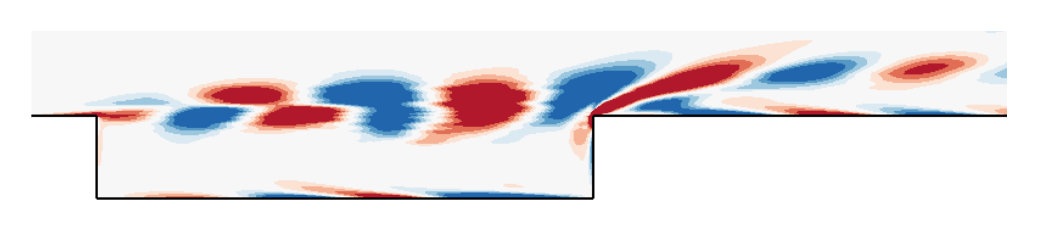}\\
$\pi$  		&\vspace{-0.1in}  \includegraphics[width=0.4\textwidth]{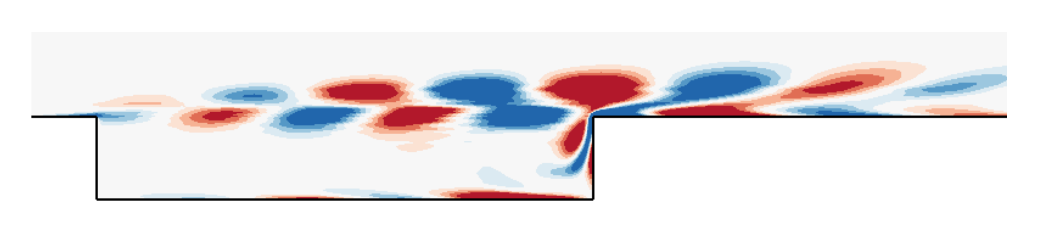}	&\vspace{-0.1in}\includegraphics[width=0.4\textwidth]{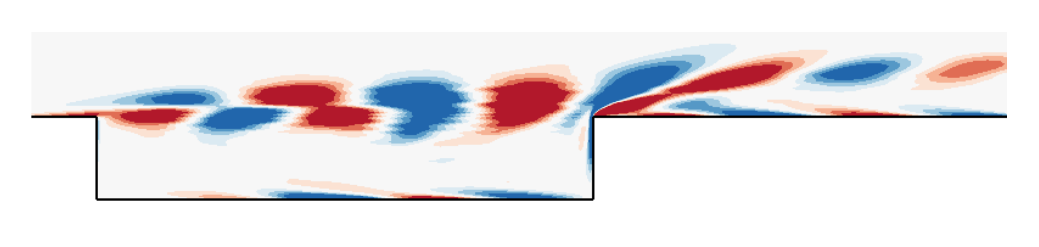}\\  \hline
\end{tabular}
}
\end{center}
\caption{Eigenvectors of the spanwise vorticity $\omega_z D/U_\infty$ at different phases $\varphi-\varphi_0$ of the Rossiter modes and the wake-modes. Here, $\varphi_0$ is a reference phase and the contour levels of $\omega_z D/U_\infty \in[-0.03,0.03]$ are used. }
\label{fig:rotate_vec}
\end{figure}

From the above discussions, we observe that the choice of the base state can significantly impact the resulting eigenmodes. In the work by Sipp and Lebedev \cite{Sipp:JFM07} on instabilities of incompressible cavity flow ($L/D=1$), they found that the use of the mean flow as base state does not predict the correct oscillation frequency revealed in the nonlinear flow, but in our present work on a long cavity ($L/D=6$), the wake-mode is captured by using the mean flow as the base state. It should be noted that Sipp and Lebedev \cite{Sipp:JFM07} focused the analysis of the shear layer mode, however, to reveal the wake-mode, the use of the mean flow as the base state is necessary. From the eigenmodes captured in this study, it should be noticed that the most-unstable mode (w-ii) has higher order frequency than the dominant frequency ($St_D=0.0572$) from the 2D nonlinear simulation. Even though the wake-mode w-i is captured using the mean flow as the base state, it is found to be stable, but near the neutral stability axis. The nonlinear interaction among modes is neglected in the global stability analysis, which likely causes the wake-modes to become unstable in the actual nonlinear flow.  It should be emphasized that the wake-mode involves a rollup of a large vortex, which constitutes a strong nonlinear dynamical process. 

\subsection{3D global stability}
\subsubsection{3D DNS}

Next, we extrude the 2D cavity setup in the spanwise direction with a periodic boundary condition to perform 3D DNS with width-to-depth ratios of $W/D=1$ and $2$. The 3D flow fields are initialized with small perturbations that a spanwise Gaussian distribution with the peak value of 5\% of free stream velocity is added to the streamwise velocity, $u$, of spanwise uniform Blasius boundary layer profiles, and analysis is performed when the flow has fully reached a periodic oscillating state. The time-history of $y-$velocity $v$ from 2D and 3D flows are then examined with Fourier analysis.  The spectra are shown in Fig.~\ref{fig:spec_comp}.  The wake-mode ($St_D=0.0572$) observed from the 2D nonlinear flow is still prominent in the 3D flow for $W/D=1$ as well as its harmonics.  The magnitude of wake-mode remains comparable between 2D flow and 3D flow with a narrow spanwise periodicity of $W/D=1$.  The iso-surface of vorticity $\omega_z D/U_\infty$ in Fig. \ref{fig:spec_comp} $(a)$ also shows the wake-mode features with close resemblance to what are observed in Fig. \ref{fig:spec}.

Although the spanwise motion $w$ emerges along the large-scale vortex roll-up in the case with $W/D=1$, as shown in Fig. \ref{fig:spec_comp} $(a)$, the magnitude of $w/U_\infty=\pm0.0015$ is almost negligible and the 3D flow maintains the characteristics from the 2D case (wake-mode).  However, as $W/D$ is increased to 2, the spectrum changes from the wake-mode behavior to the shear layer mode dominant flow. The magnitude of peak $St_D\approx0.055$ is decreased to 20\% of that in the 2D flow.  What is especially noteworthy is that one of the higher frequency modes becomes dominant but with lower amplitude than those from the other two cases (2D and 3D with $W/D=1$), suggesting reduced fluctuations being present in the flow.  The instantaneous flow field in Fig. \ref{fig:spec_comp} (b) indicates that there is no large vortex roll-up present in the flow field and an increased spanwise motion ($w/U_\infty=\pm0.1$) is captured in the aft part of the cavity. It suggests that as the spanwise width of the cavity is increased to $W/D=2$, the spanwise motion plays a critical role to modify the violent 2D wake-mode into a moderate flow behavior with a higher frequency. 
\begin{figure}[hbpt]
\centering
    \includegraphics[width=0.99\textwidth]{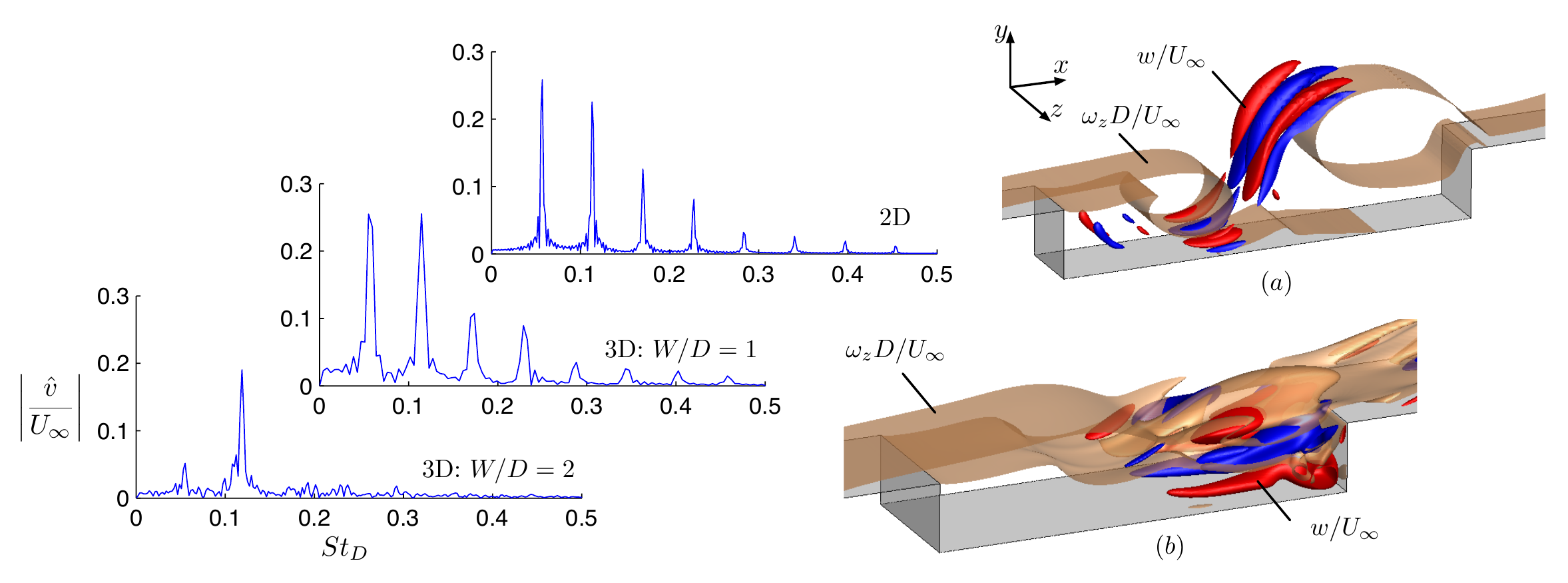}
\caption{Comparison of the spectra of the normal velocity $v/U_\infty$ from the center of the cavity $[L/2, 0]$ for the 2D as well as the 3D flows with $W/D=1$ and 2. Iso-surface of the vorticity $\omega_z D/U_\infty=-1.17$ and the spanwise velocity $w/U_\infty$ of representative instantaneous flow fields are shown on the right: $(a)$ $W/D=1$ (exhibiting wake-mode) and $(b)$ $W/D=2$ (exhibiting shear layer mode).}
\label{fig:spec_comp}   
\end{figure}

\subsubsection{3D eigenmodes for $\lambda/D=0.5-2.0$}

With the findings from the 3D DNS that the change of cavity width can alter the flow characteristics, we further perform spanwise homogeneous bi-global stability analysis to gain deeper insights on the 3D eigenmode associated with spanwise wavelengths corresponding to the range of cavity widths considered earlier. The analysis carried out in this section covers spanwise wavelengths of $\lambda/D=0.5-2.0$ with both the unstable steady state and the mean flow as the base state.  We note that the eigenspectra presented in this section only shows the values in the vicinity of the origin. 

Using the unstable steady state as the base state, the determined eigenspectra are presented in Fig. \ref{fig:3Deigenvalue_uss} with spatial structures of the least-stable 3D eigenmodes illustrated.  The eigenvalues of the leading eigenmodes are listed in Table. \ref{tab:3Deigenvalue}.  Among all wavelengths examined, the 3D eigenmodes with $\lambda/D=0.5$ are the most stable because the largest eigenvalue is $0.0092-i0.3859$ which is much smaller than those with longer wavelengths.  Amongst the dominant eigenmodes, the one with $\lambda/D=2$ is the only stationary mode with $St_D=0.0$. The iso-surfaces of the real component $w_r$ of the least-stable 3D eigenmodes are also shown in Fig. \ref{fig:3Deigenvalue_uss}. The dominant eigenmodes for $\lambda/D=0.5$ and $1.0$ are referred to as traveling mode I (Fig. \ref{fig:3Deigenvalue_uss} $(a)$), because they share similarities in the spatial structure that the perturbations are rotating around the sitting vortex in the unstable steady state. This type of 3D mode was found by Br\`es and Colonius \cite{Bres:JFM08} for shorter cavities and they attributed the mode to the centrifugal instability. The least-stable eigenmode with $\lambda/D=1.5$ is also a traveling mode with a non-zero frequency, but the spatial structures are distributed in an array along the cavity from the aft wall towards the fore of the cavity, and is referred herein as traveling mode II. A stationary 3D eigenmode becomes dominant as the spanwise wavelength $\lambda/D$ reaches 2.  Different from the traveling modes, the amplitude function of the stationary mode shows spatial oscillations at both the front and rear parts of the cavity. 
\begin{figure}[hbpt]
\begin{center}
    \includegraphics[width=1.0\textwidth]{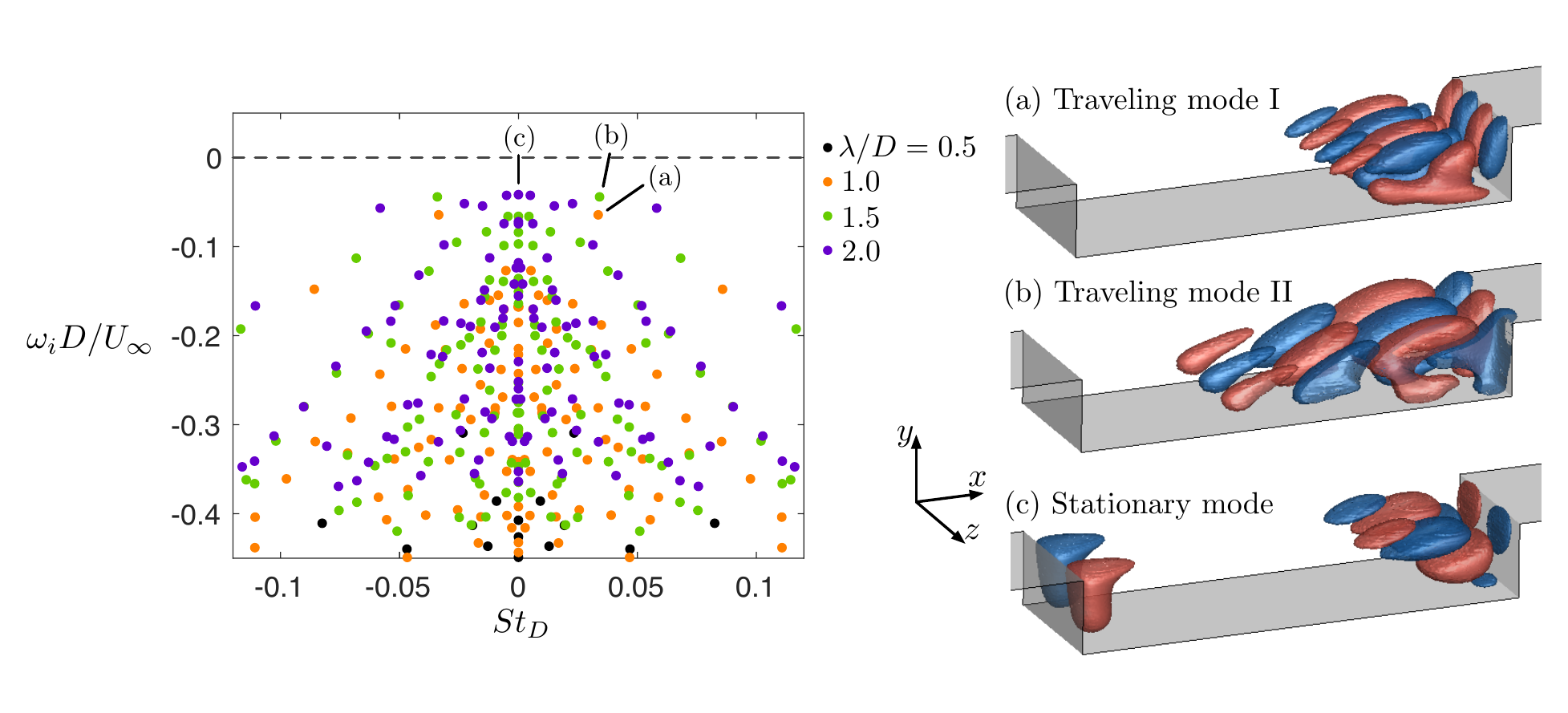}
\caption{Eigenvalues of 3D eigenmodes with spanwise wavelengths $\lambda/D=0.5-2.0$ using the unstable steady state as the base state. The least stable 3D eigenmodes are illustrated by the iso-surface of the real component $w_r$: (a) and (b) $w_r=\pm1.5\times10^{-3}$; (c) $w_r=\pm1.5\times10^{-4}$. The 3D structures  are visualized over a domain with $W/D=2$.}
\label{fig:3Deigenvalue_uss}  
\end{center}
\end{figure}

Using the mean (time-average) flow as the base state, the eigenspectra and the leading eigenvectors of 3D eigenmodes are shown in Fig. \ref{fig:3Deigenvalue_mean} and Table \ref{tab:3Deigenvalue}. Similar to the results of the least-stable eigenmodes in Fig. \ref{fig:3Deigenvalue_uss}, we find that the eigenspectra for $\lambda/D=0.5$ is much lower than the other spectra for the different spanwise wavelengths considered. It is noted that the leading eigenmodes with wavelengths $\lambda/D=1.0-2.0$ are all stationary modes. In Fig. \ref{fig:3Deigenvalue_mean}, the spatial structures of the dominant 3D eigenmodes develop across the center of the large recirculation region, and the eigenvectors are similar in the shape of structures regardless of spanwise wavelength $\lambda/D$.    

\begin{figure}[hbpt]
\begin{center}
    \includegraphics[width=1.0\textwidth]{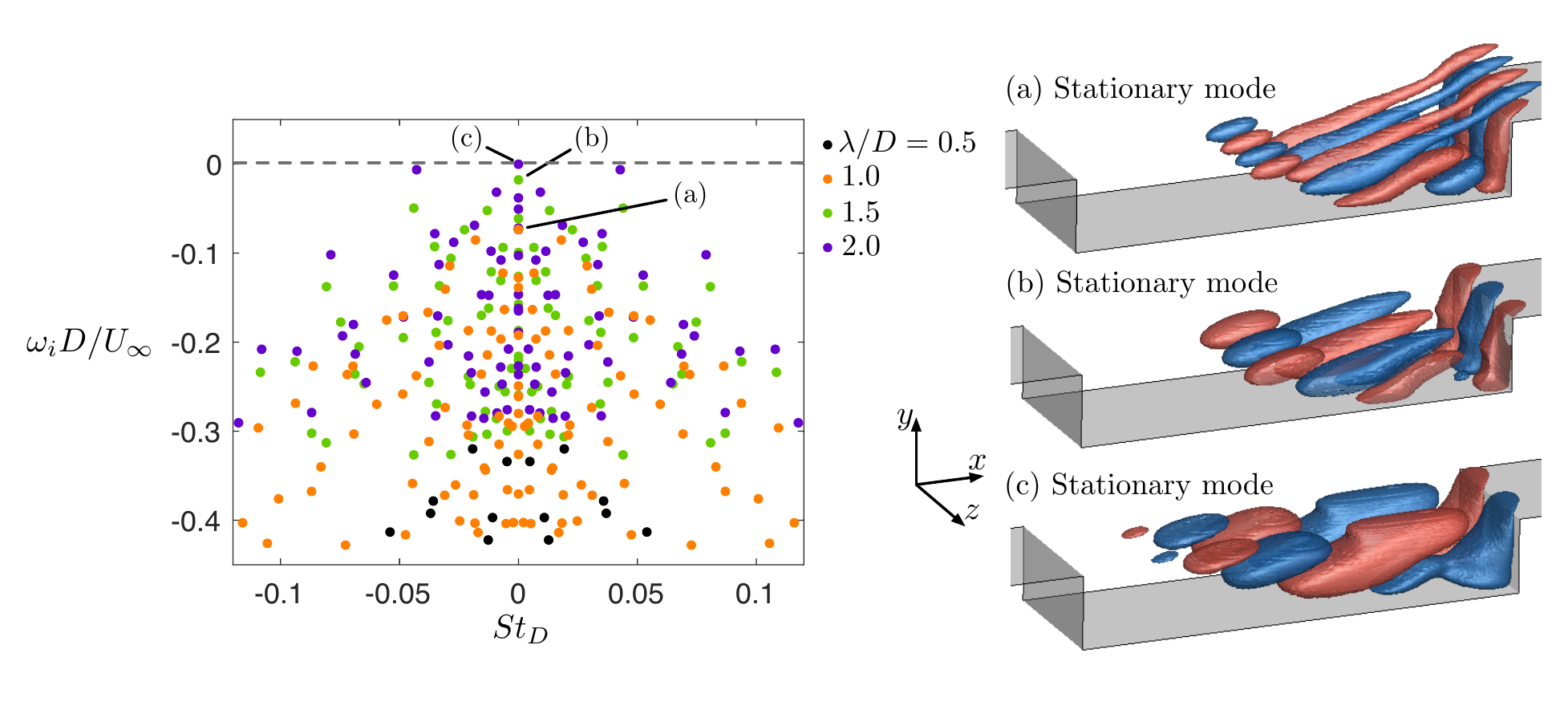}
\caption{Eigenvalues of the 3D eigenmodes with spanwise wavelengths of $\lambda/D=0.5-2.0$ using the mean (time-average) flow as the base state. The least-stable 3D eigenmodes are illustrated by the iso-surface of the real component $w_r$: (a) $w_r=\pm1.5\times10^{-3}$; (b) $w_r=\pm0.5\times10^{-4}$ and (c) $w_r=\pm0.5\times10^{-3}$. The 3D structures are visualized over a domain with $W/D=2$. }
\label{fig:3Deigenvalue_mean}  
\end{center}
\end{figure}

\begin{table}[hbpt]
\caption{Eigenvalues of the least-stable 3D eigenmodes ($\lambda/D=0.5-2.0$) using the unstable steady state and the mean flow as the base state.}
\begin{center}
{\scriptsize
\begin{tabular}{cccccc}
			&Unstable steady state			& Mean flow 					\\ \hline
$\lambda/D$ 	& $St_D+i(\omega_i D/U_\infty)$ 	& $St_D+i(\omega_i D/U_\infty)$	\\ \\
0.5			&$0.0092 - i 0.3859$				&$0.0193 - i 0.3199$				\\
1.0 			&$0.0335 - i 0.0643$				&$0.0180 - i 0.0853$				\\
1.5			&$0.0341 - i 0.0441$				&$0.0000 - i 0.0178$				\\
2.0			&$0.0000 - i 0.0416$				&$0.0000 - i 0.0003$				\\ \hline
\end{tabular}
}
\end{center}
\label{tab:3Deigenvalue}
\end{table}

Although all the captured 3D eigenmodes are stable ($\omega_i<0$) in both cases with the unstable steady state and the mean flow as the base state, relating back to the 2D global eigenmodes discussion, the spatial structures of the 2D and 3D eigenmodes overlap in the rear part of cavity.  This leads to the modal interactions and further modifies the instabilities predicted from the linear analysis, which is supported by the nonlinear simulations.  In Fig. \ref{fig:spec_comp}, we show that the spanwise motion is not apparent in the DNS and the wake-mode still dominates in the case $W/D=1$, but in the case with $W/D=2$, the spanwise motion is evident in the rear part of the cavity, in which the wake-mode is not observable. Moreover, in bi-global stability analysis, the 3D eigenmode with $\lambda/D=2.0$ is less stable than that with $\lambda/D=1.0$. Thus, the suppression of 2D wake-mode in 3D DNS can be related to the development of the 3D modes in the flow field.  As the 3D modes appear, the 2D base flow is changed such that it no longer supports the formation of the wake-mode. Hence, in experiments, the sidewall and three-dimensionality of cavity model can introduce 3D perturbations that preclude the formation of the wake-mode in experiments.

\section{Summary}
\label{sec:summary}
 
In this study, we examined the role of the spanwise instabilities on the wake-mode, whose appearance has been reported in the literature under certain conditions.  The wake-mode is captured from the 2D nonlinear simulation of a flow over a cavity of $L/D=6$ at $M_\infty=0.6$ and $Re_D=502$.  In the bi-global stability analysis with $\beta=0$, the 2D Rossiter modes and the wake-modes are uncovered by using the unstable steady state and the mean flow as the base flow, respectively. The results indicate that the choice of base state can reveal different eigenmodes from stability analysis. The use of the mean flow as the base state allows for the emergence of the wake-mode, while the use of the unstable steady state supports the appearance of shear layer modes.  The present bi-global stability analysis indicates that the use of the mean flow can identify the wake-mode which is observed in the 2D nonlinear simulation.

In the 3D DNS, the wake-mode is still prominent for narrow cavities of $W/D \lesssim 1$ but is suppressed to appear only with moderate oscillation amplitude for cavities with $W/D\gtrsim 2$. Further bi-global stability analyses are performed to uncover the 3D eigenmodes with spanwise wavelengths of $\lambda/D=0.5-2.0$. It is observed that the leading eigenmode can be traveling or stationary mode depending on $\lambda/D$.  We also find that the 3D eigenmodes can be different according to the choice of the base state. Our observations are consistent with the wake-mode not being observed in experiments, in which case the three-dimensionality from the wide cavity setup or that the sidewall triggers the spanwise instabilities, which likely prevents the wake-mode from appearing in the flow.

\bibliographystyle{spmpsci}      % mathematics and physical sciences
\bibliography{ref}   % name your BibTeX data base

\begin{thebibliography}{10}
\providecommand{\url}[1]{{#1}}
\providecommand{\urlprefix}{URL }
\expandafter\ifx\csname urlstyle\endcsname\relax
  \providecommand{\doi}[1]{DOI~\discretionary{}{}{}#1}\else
  \providecommand{\doi}{DOI~\discretionary{}{}{}\begingroup
  \urlstyle{rm}\Url}\fi

\bibitem{Akervik:PF06}
{\AA}kervik, E., Brandt, L., Henningson, D.S., H{\oe}pffner, J., Marxen, O.,
  Schlatter, P.: Steady solutions of the {N}avier-{S}tokes equations by
  selective frequency damping.
\newblock Phys. Fluids \textbf{18}, 068--102 (2006)

\bibitem{Arunajatesan:AIAA14}
Arunajatesan, S., Barone, M.F., Wagner, J.L., Casper, K.M., Beresh, S.J.: Joint
  experimental/computational inversigation into the effects of finite width on
  transonic cavity flow.
\newblock {AIAA} Paper 2014-3027 (2014)

\bibitem{Beresh:AIAA15}
Beresh, S.J., Wagner, J.L., Pruett, B.O.M., Henfling, J.F.: Supersonic flow
  over a finite-width rectangular cavity.
\newblock AIAA J. \textbf{53}(2) (2015)

\bibitem{Bres:2007}
Br\`es, G.A.: Numerical simulations of three-dimensional instabilities in
  cavity flows.
\newblock Ph.D. thesis, California Institute of Technology (2007)

\bibitem{Bres:JFM08}
Br\`es, G.A., Colonius, T.: Three-dimensional instabilities in compressible
  flow over open cavities.
\newblock J. Fluid Mech. \textbf{599}, 309--339 (2008)

\bibitem{Bres:AIAA2012}
Br\`es, G.A., Nichols, J.W., Lele, S.K., Ham, F.E.: Towards best practices for
  jet noise predictions with unstructured large eddy simulations.
\newblock 42nd {AIAA Fluid Dynamics Conference}, AIAA Paper 2012-2965, New
  Orleans (2012)

\bibitem{cattafesta2003review}
Cattafesta, L.N., Williams, D.R., Rowley, C.W., Alvi, F.S.: Review of active
  control of flow-induced cavity resonance.
\newblock 33rd AIAA Fluid Dynamics Conference, {AIAA} Paper 2003-3567 (2003)

\bibitem{Citro:JFM15}
Citro, V., Giannetti, F., Brandt, L., Luchini, P.: Linear three-dimensional
  global and asymptotic stability analysis of incompressible open cavity flow.
\newblock J. Fluid Mech. \textbf{768}, 113--140 (2015)

\bibitem{Colonius:AIAA01}
Colonius, T.: An overview of simulation, modeling, and active control of
  flow/acoustic resonance in open cavities.
\newblock 39th Aerospace Sciences Meeting and Exhibit, {AIAA} Paper 2001-0076
  (2001)

\bibitem{Ding:JCP98}
Ding, Y., Kawahara, M.: Linear stability of incompressible flow using a mixed
  finite element method.
\newblock J. Comput. Phys. \textbf{139}, 243--273 (1998)

\bibitem{George:AIAA15}
George, B., Ukeiley, L., Cattafesta, L., Taira, K.: Control of
  three-dimensional cavity flow using leading edge slot blowing.
\newblock {AIAA} Paper 2015-1059 (2015)

\bibitem{Gharib:JFM87}
Gharib, M., Roshko, A.: The effect of flow oscillations on cavity drag.
\newblock J. Fluid Mech. \textbf{177}, 501--530 (1987)

\bibitem{Krishnamurty:1956}
Krishnamurty, K.: Sound radiation from surface cutouts in high speed flow.
\newblock Ph.D. thesis, California Institute of Technology (1956)

\bibitem{Lawson:PAS11}
Lawson, S.J., Barakos, G.N.: Review of numerical simulations for high-speed,
  turbulent cavity flows.
\newblock Prog. Aero. Sci. \textbf{47}, 186--216 (2011)

\bibitem{Arpack:96}
Lehoucq, R., Maschhoff, K., Sorensen, D., Yang, C.: {ARPACK} software
  (1996-2007).
\newblock \urlprefix\url{http://www.caam.rice.edu/software/ARPACK/}

\bibitem{Maull:JFM63}
Maull, D.J., East, L.F.: Three-dimensional flow in cavities.
\newblock J. Fluid Mech. \textbf{16}, 620--632 (1963)

\bibitem{Munday:inhouse16}
Munday, P., Taira, K.: private communication  (2016)

\bibitem{Ramanan:PF94}
Ramanan, N., Homsy, G.M.: Linear stability of lid-driven cavity flow.
\newblock Phys. Fluids \textbf{6}, 2690--2701 (1994)

\bibitem{Rockwell:JFE78}
Rockwell, D., Naudascher, E.: Review-self-sustaining oscillations of flow past
  cavities.
\newblock J. Fluids Eng. \textbf{100}, 152--165 (1978)

\bibitem{Rossiter:ARCRM64}
Rossiter, J.E.: Wind-tunnel experiments on the flow over rectangular cavities
  at subsonic and transonic speeds.
\newblock Tech. Rep. 3438, Aeronautical Research Council Reports and Memoranda
  (1964)

\bibitem{Rowley:JFM02}
Rowley, C.W., Colonius, T., Basu, A.J.: On self-sustained oscillations in
  two-dimensional compressible flow over rectangular cavities.
\newblock J. Fluid Mech. \textbf{455}, 315--346 (2002)

\bibitem{Schmid01}
Schmid, P.J., Henningson, D.S.: Stability and transition in shear flows.
\newblock Springer (2001)

\bibitem{Sipp:JFM07}
Sipp, D., Lebedev, A.: Global stability of base and mean flows: a general
  approach and its applications to cylinder and open cavity flows.
\newblock J. Fluid Mech. \textbf{593}, 333--358 (2007)

\bibitem{Sun:AIAA2014}
Sun, Y., Nair, A.G., Taira, K., Cattafesta, L.N., Br\`es, G.A., Ukeiley, L.S.:
  Numerical simulation of subsonic and transonic open-cavity flows.
\newblock {7th AIAA Theoretical Fluid Mechanics Conference}, {AIAA} Paper
  2014-3092 (2014)

\bibitem{Sun:JFM16}
Sun, Y., Taira, K., Cattafesta, L., Ukeiley, L.: Global stability analysis of
  compressible open-cavity flows.
\newblock (in preparation)  (2016)

\bibitem{Sun:AIAA16}
Sun, Y., Zhang, Y., Taira, K., Cattafesta, L., George, B., Ukeiley, L.: Width
  and sidewall effects on high speed cavity flows.
\newblock 54th AIAA Aerospace Sciences Meeting, {AIAA} Paper 2016-1343 (2016)

\bibitem{Theofilis:PAS03}
Theofilis, V.: Advances in global linear instability analysis of nonparallel
  and three-dimensional flows.
\newblock Prog. Aero. Sci. \textbf{39}, 249--315 (2003)

\bibitem{Theofilis:ARFM11}
Theofilis, V.: Global linear instability.
\newblock Annu. Rev. Fluid Mech. \textbf{43}, 319--352 (2011)

\bibitem{Theofilis:AIAA04}
Theofilis, V., Colonius, T.: Three-dimensional instabilities of compressible
  flow over open cavities: direct solution of the biglobal eigenvalue problem.
\newblock 34th Fluid Dynamics Conference and Exhibit, {AIAA} Paper 2004-2544
  (2004)

\bibitem{Theofilis:JFM04}
Theofilis, V., Duck, P.W., Owen, J.: Viscous linear stability analysis of
  rectangular duct and cavity flows.
\newblock J. Fluid Mech. \textbf{505}, 249--286 (2004)

\bibitem{Vicente:JFM14}
de~Vicente, J., Basley, J., Meseguer-Garrido, F., Soria, J., Theofilis, V.:
  Three-dimensional instabilities over a rectangular open cavity: from linear
  stability analysis to experimentation.
\newblock J. Fluid Mech. \textbf{748}, 189--220 (2014)

\bibitem{Yamouni:JFM13}
Yamouni, S., Sipp, D., Jacquin, L.: Interaction between feedback aeroacoustic
  and acoustic resonance mechanisms in a cavity flow: a global stability
  analysis.
\newblock Journal of Fluid Mechanics \textbf{717}, 134--165 (2013)

\bibitem{Zhang:AIAA06}
Zhang, K., Naguib, A.M.: Dispersion relation and mode selectivity in
  low-{M}ach-number cavity flows.
\newblock pp. 1072--1086. 36th AIAA Fluid Dynamics Conference (2006)

\bibitem{Zhang:EF11}
Zhang, K., Naguib, A.M.: Effect of finite cavity width on flow oscillation in a
  low-{M}ach-number cavity flow.
\newblock Experiments in Fluids \textbf{51}(5), 1209--1229 (2011)

\bibitem{Zhang:AIAA15}
Zhang, Y., Sun, Y., Arora, N., Cattafesta, L., Taira, K., Ukeiley, L.:
  Suppression of cavity oscillations via three-dimensional steady blowing.
\newblock {AIAA} Paper 2015-3219 (2015)

\end{thebibliography}

\end{document}